\documentclass[twoside]{article}
\usepackage{fleqn}
\usepackage{espcrc2}
\usepackage{graphicx}
\title{Properties of tangential and radial angles of muons in EAS
{\thanks{Presented at XIIth ISHVECRI, CERN 2002}}}

\author{J. Zabierowski\address{Soltan Institute for Nuclear Studies, Cosmic Ray Phys. Dept., \\
        P.O.Box 447, PL-90-950 Lodz, Poland}
	\thanks{e-mail: janzab@zpk.u.lodz.pl},
	K. Daumiller\address{Institut f\"ur Experimentelle Kernphysik, Universit\"at Karlsruhe,\\
         P.O.Box 3640, D-76021 Karlsruhe, Germany}
        and
       P. Doll\address{Institut f\"ur Kernphysik, Forschungszentrum Karlsruhe, \\
             P.O. Box 3640, D-76021 Karlsruhe, Germany}}

\begin{document}

\begin{abstract}
Tangential and radial angles of muons in EAS, a useful concept in investigation
of the muon production height, can be used also for the investigation of the
muon momenta. A parameter $\zeta$, being a combination of tangential and
radial angles, is introduced and its possible applications in investigation
of muons in showers are presented.
\end{abstract}

\maketitle

\section{Introduction}

EAS experiments, combining field detector arrays - being capable in measuring
the shower direction - with muon tracking detectors, can be used for determination
of the mean muon production height (MPH), a parameter sensitive to the primary
mass. Muon tracks in air shower are, in general, not coplanar with the shower
axis, which have led to the concept of tangential and radial angles \cite{bern}.
However, as it will be shown in the following, these two independent observables
may provide a possibility to investigate the muon momentum components in showers.
A good example of the EAS experiment, where the ideas presented in this paper
can be used is KASCADE \cite{klages} with its recently introduced large Muon
Tracking Detector \cite{doll}. We are using data simulated with CORSIKA \cite{heck}
ver.5.644 and QGSJet hadronic interaction model for all discussions and examples
given below.

\section{Tangential and radial angles}

\begin{figure}
\begin{center}
\includegraphics[width=15pc]{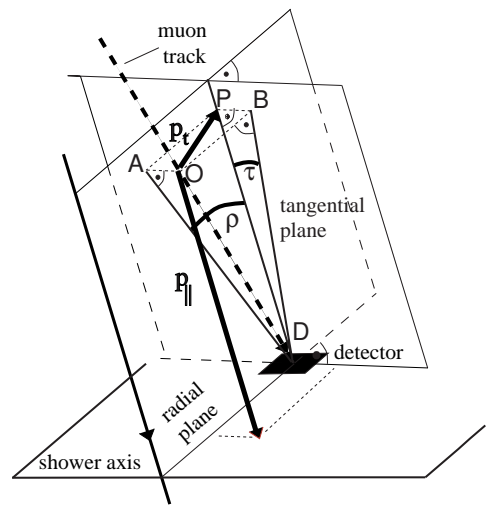}
\vspace{-0.5cm}
\caption{Definitions of tangential and radial angles.}
\label{angles}
\end{center}
\end{figure}

As it is shown in Fig. \ref{angles}, shower direction and the location of the
muon detector define two perpendicular planes: {\it tangential} and
{\it radial} ones.

 Tangential angle $\tau$ and radial angle $\rho$ are angles between
the shower direction and the orthogonal projections of the muon track onto the
tangential and radial planes, respectively.

\begin{figure}[ht]
\begin{center}
\vspace{-10mm}
\includegraphics[width=16pc]{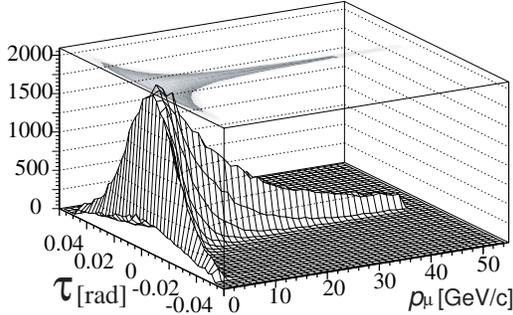}
\vspace{-5mm}
\caption{Dependence of the tangential angle on muon momentum.}
\label{taucont}
\end{center}
\end{figure}

The measured value of $\tau$ reflects the amount of muon scattering
in the atmosphere and any relevant absorber/detector material, together with
a possible displacement of the muon production place from the shower axis. Its distribution
(Fig. \ref{taucont}) is symmetrical around zero and gets narrower with the increase
of the muon momentum, what one could expect from the quantity related to the
multiple scattering.

\begin{figure}[ht]
\begin{center}
\vspace{-25mm}
\includegraphics[width=16pc]{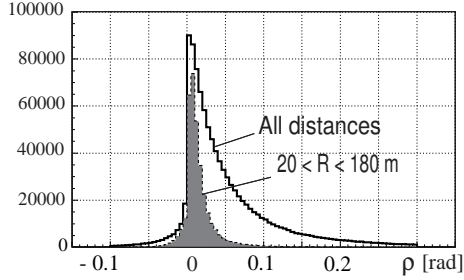}
\vspace{-1cm}
\caption{Radial angle distribution.}
\label{rad}
\end{center}
\end{figure}

A typical distribution of the radial angle is shown in Fig. \ref{rad}. Its value
depends on the muon parent's (meson's) transverse momentum and the distance
of the muon to the shower axis at the observation level. In real experimental
conditions muons are sampled within certain range of distances from the core.
The shaded distribution is shown for a typical distance 20 m - 180 m from the
core. The large $\rho$ values are removed.

\section{Parameter $\zeta$}

As it is seen from Fig. \ref{angles}, using simple trigonometry and substituting
tangent for an angle, what is correct within 5$\%$ error up to $0.4 rad$,
one can write the following expressions for $\tau$ and $\rho$:
\large
\begin{center}
\( \tau \cong \frac{PB}{PD}=\frac{p_{t}}{p_{\Vert }}\times \sin \angle POB \)

\( \rho \cong \frac{PA}{PD}=\frac{p_{t}}{p\Vert }\times \cos \angle POB \)
\end{center}
\normalsize
Let us define the new parameter $\zeta$ as follows:
\large
\begin{center}
\( \varsigma \equiv \sqrt{\tau ^{2}+\rho ^{2}}=\frac{p_{t}}{p\Vert } \) ;
\end{center}
\normalsize

which is valid for $\tau $ $\leq$ 0.4 rad, and $\rho$ $\leq$ 0.4 rad.

So, measuring in an experiment angles of a shower and of muons and calculating $\rho$
and $\tau$, by means of $\zeta$ one gets the possibility to investigate the muon
momentum space. The precision of $\zeta$ depends on the precision of angular
measurements in the air shower, and, in particular, on the precision of muon tracking.
It is known, that the tracking accuracy improves when the high energy muons only are
considered. Therefore, one would like to have as large content of high energy muons in
the analyzed data sample as possible.
\begin{figure}[ht]
\begin{center}
\vspace{-10mm}
\includegraphics[width=14pc]{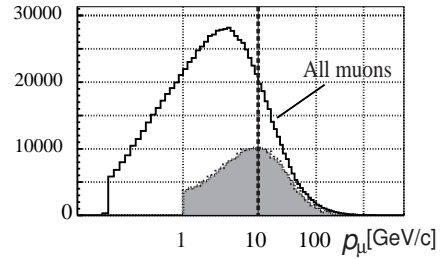}
\vspace{-20mm}
\caption{Muon momentum distribution in showers initiated by proton primaries with the
energy $E_0$=$10^{15}$ eV. The shaded area shows the effect of the experimental
threshold 1GeV/c and the limited distance to the shower core: 20 - 180 m.}
\label{mendistr}
\end{center}
\end{figure}

A typical muon momentum distribution in showers initiated by proton primaries with
$E_0$=$10^{15}$ eV is shown in Fig. \ref{mendistr}. In all simulated showers (solid
line) $\approx$ $30\%$ are muons with momentum larger than 10 GeV/c. In real
experimental conditions, where a threshold, eg. 1~GeV/c, is present and muons are
registered in a restricted range of distances to the shower core, one gets the shaded
distribution in Fig. \ref{mendistr}. In this sample, with significantly reduced
statistics, there is already $\approx$ 45 $\%$ of muons with $p_{\mu}$ $\geq$ 10~GeV/c.
However, one would like to increase high energy muon content in the data sample even
more, and for this purpose the $\zeta$ parameter can be used.

\begin{figure}[ht]
\begin{center}
\vspace*{-12mm}
\includegraphics[width=14pc]{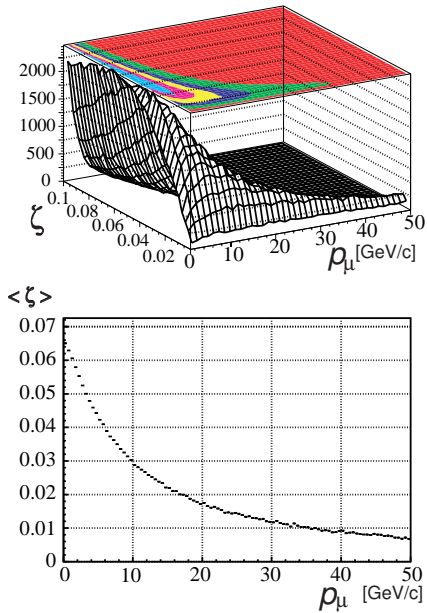}
\vspace{-8mm}
\caption{Dependence of $\zeta$ on the muon momentum. In the lower panel a mean value of
$\zeta$ is plotted.}
\label{zetamom}
\end{center}
\end{figure}

Indeed, as seen from Fig. \ref{zetamom}, $\zeta$ shows strong dependence on the muon
momentum. While at low momenta the full spread of $\zeta$ values is observed, its
distribution becomes more confined to the low values with the increase of the muon
momentum.

The mean value of $\zeta$ decreases with momentum exponentially, what is shown in the
lower panel of Fig. \ref{zetamom}. So, it is possible to use in the analysis the cut
on the $\zeta$ value and, in this way, to increase the high energy muon content in the
analyzed sample.

\begin{figure}[ht]
\begin{center}
\vspace*{-10mm}
\includegraphics[width=15pc]{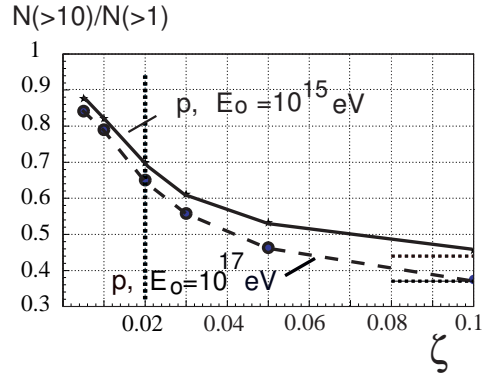}
\vspace{-10mm}
\caption{Relative content of muons with momentum larger than 10 GeV/c in the whole data
sample (muons above 1 GeV/c) as a function of applied $\zeta$ cut.}
\label{ratio}
\end{center}
\end{figure}

In Fig. \ref{ratio} the ratio of muons with momentum larger than 10 GeV/c to all muons
in a sample (above the threshold of 1 GeV/c) as a function of the applied cut on
$\zeta$ value is plotted for two primary proton energies. Only muons in the
distance between 20 and 180 m to the shower core were considered. It is seen that
choosing for the analysis muons e.g. with $\zeta$ $<$ 0.02 results in significant
increase of the high energy muon content in the data sample.

Another application of the $\zeta$ parameter is rather straightforward. Its value is
nothing else but a tangent of the muon angle versus the shower axis. So, knowing the
distance to the shower core a simple triangulation allows to reproduce the muon
production height. It is, off course, assumed that muons stem from the shower axis,
what is generally not true. But, it turns out, that such approximation works well.

\begin{figure}[ht]
\begin{center}
\includegraphics[width=15pc]{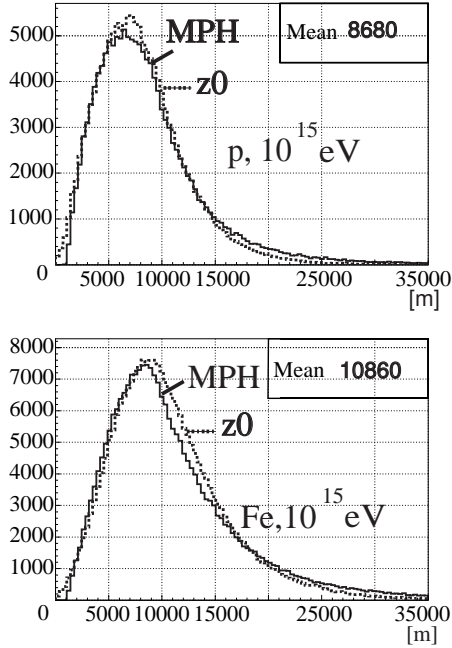}
\vspace{-5mm}
\caption{Comparison of the true ($z_0$) and reconstructed, using parameter $\zeta$, (MPH)
muon production height distributions for two primaries, proton and iron, of
$E_0$=$10^{15}$ eV. Data sample was enriched in high energy muons by selecting muons
with $\zeta$ $<$ 0.02.}
\label{mph}
\end{center}
\end{figure}
In Fig. \ref{mph} the true muon production height distributions ($z_0$) are compared
with the calculated ones (MPH), using the just mentionned procedure, for two primary
species, H and Fe with energy $E_0$=$10^{15}$ eV. In these calculations a cut on
$\zeta$ $<$ 0.02 was applied. One sees a good agreement of both
curves. The mean production height values are separated by more than 2 kilometers for
proton and iron. So, one can use this procedure of calculating mean muon production
height for the primary mass determination.

Parameter $\zeta$ can be used for investigation of muon rapidities in showers. Substituting
momentum for the total energy, which is allowed above 1~GeV without signifcant error, one can
 express rapidity $y$ and pseudorapidity $\eta$ using parameter $\zeta$ as follows:

\begin{center}
\large
\( y=\frac{1}{2}\ln \frac{E+p_{\parallel }}{E-p_{\parallel }} \)\( \approx \frac{1}{2}\ln \frac{(\sqrt{\zeta ^{2}+1})+1}{(\sqrt{\zeta ^{2}+1})-1} \)

\vspace{5mm}

\( \eta =\ln \frac{2\times p_{\parallel }}{p_{t}} \)\( \approx - \)\( \ln \frac{\zeta }{2} \)

\normalsize
\end{center}

As an example in Fig. \ref{eta} pseudorapidiy distribution of muons in showers initiated
by proton primary with the energy $E_0$=$10^{16}$ eV is shown. Only muons with momentum larger than
1~GeV/c are considered. Dotted line shows all pseudorapidity values for all simulated muons.
Solid line corresponds to the case, where $\eta$ can be calculated using parameter $\zeta$, i.e. for
muons with absolute values of $\tau$ and $\rho$ less than 0.4 radian. Dashed line corresponds to the
case, where the distance of muons to the shower core is in the range of 20~-~180~m. The large values,
being the most interesting ones,
are cut away due to 20 m minimum distance, so it is important to come with measurements as close to the
core as possible.
\begin{center}
\vspace{-25mm}
\begin{figure}[ht]
\includegraphics[width=17pc]{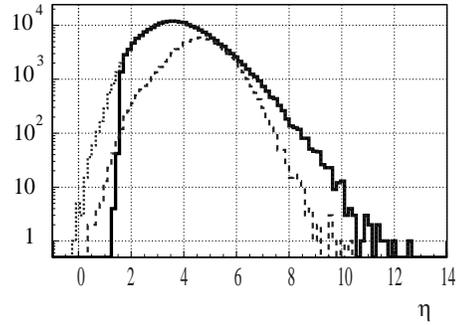}
\vspace{-1.3cm}
\caption{Muon pseudorapidity distribution. For explanations see text.}
\label{eta}
\end{figure}
\end{center}
\vspace{-1cm}
The above example shows, that by means of $\zeta$ it is possible to investigate rapidities of muons in
showers. In general, measuring directional parameters of muons in showers by means of the $\zeta$
parameter one can investigate muon momentum space.

The authors acknowledge very much support for this work obtained from Polish State
Committee for Scientific Research (grant
No. 5 P03B 133 20) and from the German Federal Ministry of Research (05 CU1VK1/9, WTZ POL 99/005).

\end{document}